\documentstyle[12pt]{article}

\newcommand{\F}{\noindent}

\newcommand{\q}{\qquad}
\newcommand{\SP}{\smallskip}
\newcommand{\MP}{\medskip}
\newcommand{\BP}{\bigskip}

\newcommand{\beq}{\begin{eqnarray}}
\newcommand{\ene}{\end{eqnarray}}

\newcommand{\HH}{{\cal H}}
\newcommand{\UU}{{\cal U}}
\newcommand{\OO}{{\cal O}}
\newcommand{\la}{{\lambda}}

\newcommand{\Ltnn}{{L^2(R^{3n})}}

\newcommand{\ve}{\vert}

\begin{document}

\Large

\begin{center}
{\bf Theory of Local Times}
\end{center}

\BP

\normalsize

\begin{center}
 Hitoshi Kitada

Department of Mathematical Sciences, University of
Tokyo

Komaba, Meguro, Tokyo 153, Japan
\end{center}

\SP

\begin{center}
(published in Il Nuovo Cimento -- Vol. 109 B, N. 3 -- March 1994 --
 pp. 281-302)
\end{center}
\centerline{\ }

\MP

\leftskip=24pt
\rightskip=24pt

\small

\F
{\bf Summary.} --- A model of a stationary universe is proposed.
In this framework, time is defined as a local and quantum-mechanical notion in
 the sense that it is defined for each local and quantum-mechanical system
 consisting of finite number of particles.
The total universe consisting of infinite number of particles has no time
 associated.
It is a stationary bound state of the total Hamiltonian of infinite degrees
 of freedom.
The quantum mechanics and the theory of general relativity are consistently
 united in this context if one uses this notion of local and
quantum-mechanical time.
As one of the consequences, the Einstein-Podolsky-Rosen paradox is resolved.
The Hubble   red-shift is explained as a consequence of general relativity
which is consistent with quantum mechanics.
This does not require us to argue on the beginning nor the end of the universe.
The universe just exists without time.
\MP

\noindent
PACS O3.65.Bz - Foundations, theory of measurement, miscellaneous theories.

\leftskip=0pt
\rightskip=0pt

\normalsize

\vskip 20pt

\BP
\F
{\bf 1. -- Introduction.}
\normalsize

\vskip 12pt

As stated in the abstract, the main theme of the paper is to present one
 possible consistent unification of quantum mechanics and general relativity.
This is stated intentionally with anticipating the naive refutation that the
Euclidean geometry which quantum mechanics follows and the non-flat Riemannian
geometry which relativity follows can never be united consistently.

Our trick of the consistent unification of these two theories is to adopt a
ten-dimensional vector bundle $X \times R^6$ (the reason $R^6$ is adopted
instead of $R^4$ will be touched below) as the total physics space, where the
base space $X$ and the fibre $R^6$ are mutually orthogonal.
    Quantum mechanics is set on the Euclidean space $R^6$ and relativity
 theory on the curved Riemannian space $X$.
Each point $(t,x)\in X$ is correlated to the {\it centre} of mass of the local
system consisting of finite number of (quantum-mechanical) particles, and these
 centres of mass are considered as the {\it classical} particles.
These classical particles are regarded as moving following general
relativity in the Riemannian manifold $X$ on the one hand, and the particles
inside the local systems are regarded as moving following quantum mechanics on
the other hand.

In this sense each point $(t,x)$ of the base Riemann space $X$ of the vector
bundle $X\times R^6$ corresponds to the local system consisting of finite
number of particles which follow quantum mechanics in each fibre $R^6$.

Because the fibre $R^6$ where quantum mechanics holds and the base space $X$
 where     relativity holds are mutually orthogonal in the total space
$X\times R^6$, it can be shown that the principles of quantum mechanics and
the principles of general relativity are united consistently in this
formulation, with the aid of the notion of quantum-mechanical local time
introduced after sect. 2 below.

The reader may ask: There are many sublocal systems in a local system
$(H, \HH)$, and the centres of mass of these sublocal systems follow classical
relativity.
But inside the local system $(H,\HH)$, quantum mechanics holds.
Then what mechanics do these sublocal systems follow?
The answer is

\MP

\ \ \
These  sublocal systems follow     classical relativity {\it as far as} the
 observer observes them in accordance with {\it the observer's} own time, but
 if the {\it time} of the system $(H,\HH)$ itself is adopted as the reference
 time, then the local system $(H,\HH)$ follows quantum mechanics.

\MP

Now here enters the notion of local time, which is the main ingredient of our
 consistent unification of the quantum and relativity theories.

Our starting point is the recognition that the time $t$ is given by the ratio
 $|x|/|v|$ of the position $x$ and the motion (velocity) $v$.
This standpoint is fully explained in sect. 2.
This formulation is justified by the result (theorem l) of the many-particle
 scattering theory.
Because of this definition of time, the Euclidean space of quantum mechanics
 becomes $R^6$ of $(x,v)$ instead of $R^4$ of $(t,x)$, and the usual
four-dimensional structure $R^4$ is recovered as an approximation through the
 uncertainty principle stated in sect. 5 after {\it definition} 3.

Like this, quantum mechanics controls the mechanics inside each local system,
 and the relative motion of the centres of mass of these local systems follows
 general relativity.
In actual observations or experiments, the observer's time must be transformed
into the local time of the observed local system.
The rules of this transformation are given by the covariance principle and the
equivalence principle of general relativity.
An actual example of this kind of explanation of the relativistic quantum
 phenomena is given in sect. 9.

As a consequence, the EPR paradox is explained without contradiction, and
 Hubble's red-shift is explained even though our model is constructed on the
 basis of a stationary universe.
As another result, the dark matter is unnecessary in our theory to explain the
stability of galaxies, clusters of galaxies, and so on.
\vskip 18pt

\F
{\bf 2. -- What is time?}
\normalsize

\vskip 8pt

This question seems to have a history as long as that of mankind itself.
In the literature of physics, in the seventeenth century time was defined by
 Newton as a kind of absolute notion.
In the first decade of the twentieth century, a reflection on the notion of
 time was given by Einstein in his theory of special relativity.
Later, Einstein gave a more profound reflection on space-time in his theory of
general relativity.
Even though these reflections required us to make a reconstruction of our
 notion of space-time on a very deep level, time together with space is still
 considered to give a reference frame according to which one measures the
physical quantities like positions, motions, velocities, and so on:
 Matter determines the space-time structure of the universe according to the
theory of general relativity.
 Nonetheless, the space-time remains as a frame according to which the
position of each matter is determined.

What we propose in this paper is a different point of view on the notion of
 time based on the following simple but overlooked observation:
 time does not appear until it is measured by some equipments, {\it clocks}.
We need tools to know the time.
This situation is different from the perception of the positions and motions,
 which are recognized directly by our senses.
Even when we use some tools like a rule to measure the length of a thing, what
 we actually do is to {\it see} which scales of the rule coincide with both
 sides of the thing.
The fundamental recognition done here is the perception of this coincidence,
 which cannot be done without our senses.
The positions and motions are recognized through our senses in this sense.
On the contrary, if one reflects the measurement procedure of time by clocks,
  one notices that he or she compares several motions or movements of matters,
 and takes the quotient of position and velocity.
In fact, the usual (analog) clock measures time by the motion of its hands.
 We look at the hands, and recognize that one second passes if the second hand
 $\langle\langle$moves$\rangle\rangle$ one
 $\langle\langle$scale$\rangle\rangle$.
We do not measure time directly by our senses, but we know time
 by perceiving the positions and motions of the hands of clocks.
In this sense time is neither a quantity nor a frame given a priori.
What exists first are the positions and movements of the matters relative
 to our own position.
The perception of the positions and motions indicates an introduction of
 the common parameter in each system of matters consisting of {\it finite}
 number of particles.
This parameter is called {\it time} and it is a local notion by nature.
This recognition is our starting point.

In the twentieth century several theories of the universe have been
 investigated.
Most of these are related with the theory of (general) relativity.
Many trials have been made to $\langle\langle$quantize$\rangle\rangle$
 the theory of relativity.
But it seems that none of them can be said to have succeeded.
(See Ashtekar's Introduction of Ashtekar and Stachel [1] and Streater's
 paper in Brown-Harr\'e [2] for the history of many trials including
 the recent ones.
See also Glimm-Jaffe [3] for the point of view that quantum field theory
 is successful to a {\it certain} extent.)
 One category of such theories is represented by the stationary theories
 of the universe.
Another category consists of non-stationary ones.
The typical theory of the latter category is the so-called
 $\langle\langle$big-bang$\rangle\rangle$ theory.
Tracing in this direction, Hawking introduced, for instance, imaginary time.

In this article we introduce one stationary model of universe.
This enables us to define the notion of (local) times.
According to our theory, there exists the total universe which has no time
 associated.
It is a stationary bound state of the total Hamiltonian with infinite
 degrees of freedom.
 Our theory is quantized from the beginning in this context.
 Relative coordinates and momenta are postulated to exist according to
 our reflection of the measuring procedure of time stated above.

{}From these postulates we define {\it time} as a local
 (or $\langle\langle$glocal$\rangle\rangle$) and quantum-mechanical quantity.
Our theory may also be called a quantization of time in this sense.
We further show that our theory is consistent with the theory of
 general relativity.
According to the so-called EPR paradox (see Redhead [4], for instance),
 quantum mechanics and the theory of relativity cannot be consistently
 united, because the former denies the local causality which is a
 consequence of the latter, as far as we suppose that quantum mechanics
 is complete in the sense described below in the argument of the EPR
 paradox in sect. 8.
In this sense the unification of these two theories is impossible on the
 same level or on the same $\langle\langle$plane$\rangle\rangle$.
In our unification of these theories, we
 $\langle\langle$orthogonalize$\rangle\rangle$, so to speak, these
 two theories, or set these theories on mutually
 $\langle\langle$orthogonal planes$\rangle\rangle$, with the
 introduction of local times.
We do not try to quantize the theory of relativity itself,
 as it has been tried in many theories of quantum fields.
We just leave the theory of general relativity as a classical theory.
What we actually do is to show that our theory of local
 quantum-mechanical times is consistent with classical relativity.

Our theory is basically the non-relativistic quantum mechanics
 $\langle\langle$orthogonalized$\rangle\rangle$ to general relativity.
The actual way of this orthogonalization is shown in sect. 6.
We take the standpoint that {\it nature follows quantum mechanics
 intrinsically, and the relativistic effects only appear associated
 with observation.}
Namely, gravitation is nothing but an outlook; it is no intrinsic
 nature of the universe according to our theory.
Note that this standpoint is different from the standpoint of quantum
 gravity being discussed recently, {\it e.g.}, in Ashtekar and
 Stachel [1], where it is implicitly supposed that the quantum and
 gravitational aspects of the universe are possible to be discussed
 on the same $\langle\langle$plane$\rangle\rangle$ or on the same level.
Our theory differs from this traditional standpoint.

The point of our theory of orthogonalization of quantum mechanics
 and relativity theory is summarized as follows:

\medskip

\ \ \
The {\it observer's} coordinate system $(t^0,x^0)$ of $R^4$ is
 {\it independent of} the coordinate system $(t^1,x^1)$ of the
 {\it observed} system, according to our definition of local times
 $t^1,t^0$ (sect. 5).
Thus the observed system can follow quantum mechanics {\it with respect to}
 the coordinate system $(t^1,x^1)$, as well as it can follow classical
 relativity {\it with respect to} the coordinate system $(t^0, x^0)$.
In this sense, any local system can follow classical relativity
 {\it when observed}, and at the same time, it can follow quantum
 mechanics {\it inwardly} or {\it intrinsically}.

\medskip

This is our key of the consistent unification of quantum mechanics
 and general relativity.

Accordingly, any local system behaves in classical-mechanical way
 as well as in quantum-mechanical way, depending on the choice of
 coordinate systems.
For the relation between these two looks of the behaviour of local
 systems, we make the following fundamental assumption:

\medskip

\ \ \
{\it The observer can see only the motions of} centres {\it of mass of
 local systems, and these motions are observed following classical
 relativity.
The quantum effects} inside {\it the local systems are unobservable
 directly, but can be deduced from these classical observations of
 the sublocal systems.}

\medskip

The method of deduction of the internal quantum mechanics from the
 classical observations which we propose is discussed in sect. 9
 as a procedure of observation.

The unification of quantum and relativity theories in the above
form means that we abandon the {\it relativistic} quantum field theory.
A positive reason which supports this abandonment is that there seems
to be only a trivial model in the axiomatic quantum field theory in actual
 four-dimensional space-time (see Streater's paper in [2]).
Owing to this abandonment of quantum field theory, we are free from such
 problems as renormalizations, divergence problems, and so on, which
 constitute the fundamental difficulties of quantum field theories.
The classical fields are not quantized {\it relativistically} in our theory.
They are left as classical notions.
The quantum fields are auxiliary tools, which will be introduced to
 treat the creation and annihilation phenomena (see conjecture (2) below).
The quantum fields are confined to the explanation of the non-relativistic
 quantum phenomena.

Summarizing, the most important result of the paper is that there exists
at least one consistent theory which unites the {\it non}-relativistic
quantum mechanics and general relativity.
(As an example of other trials in this direction, we refer to
 Prugove\v cki [5], where a geometro-stochastic approach is adopted
 for the problem of the unification of quantum theory and general
 relativity.)
Even in our approach, we can give an explanation of the so-called
{\it relativistic} quantum phenomena.
These phenomena will be explained as the consequences of the observation
 activities: the actual experimental data are different from their
{\it true} quantum-mechanical values because of the relativistic effects
 of observations, and the observer has to calculate the true values from
 his experimental data through some relativistic considerations.
This point will be discussed in sect. 9 as regards observation.
The necessity of considering high-energy particle physics is not strong
in our context, since the universe does not begin via something like
the big bang.
However, some of the high-energy phenomena related with relativistic
quantum mechanics can be explained in our context.
Hopefully, a full explanation of these phenomena would be given in the future.

\vskip 18pt

\F
{\bf 3. -- Global axiom.}

\normalsize

\vskip 8pt

As a foundation of the theory of times, we adopt a model
of stationary universe.
Differently from the usual stationary theories where the time of
the universe is assumed to exist a priori, our universe $\phi$, which
 is an element of a Hilbert space $\UU$, has no time associated.
It is assumed as a stationary bound state of a Hamiltonian $H$ of infinite
 degrees of freedom.
In this sense our universe $\phi$ is quantized from the beginning:

\BP

{\it Axiom} 1.\
There exists a (separable) Hilbert space of (possible) universes
\beq
\UU=\{\phi\}=\bigoplus_{n=0}^\infty \left(\bigoplus_{\ell=0}^\infty
\HH^n  \right) \q (\HH^n=\underbrace{\HH\otimes\cdots\otimes\HH}_{n\
\mbox{\scriptsize{factors}}})
\ene
with $\HH$ being a separable Hilbert space,   a set of self-adjoint operators
(observables) $\OO=\{ A\}$ of the form $A\phi=(A_{n\ell}\phi_{n\ell})$ for
$\phi=(\phi_{n\ell})\in\UU$, and a self-adjoint operator $H\in\OO$ in $\UU$
such that for some $\phi\in \UU-\{0\}$ and $\la\in R^1$
\beq
H\phi=\la \phi
\ene
in the following sense:
There exists an infinite matrix $(\la_{n\ell})$ of real numbers such that
$H_{n\ell}\phi_{n\ell}=\la_{n\ell}\phi_{n\ell}$ for each $n\ge1$, $\ell\ge 0$
and $\la_{n\ell_n}\to\la$ as $n\to\infty$ along any $\ell_n$ such that
$F_n^{\ell_n}\subset F_{n+1}^{\ell_{n+1}}$. Here $F_n$ is a finite subset of
${\mbox{\bf N}}=\{1,2,\cdots\}$ with $\sharp(F_n)=$ (the number of elements in
$F_n)=n$ and
$\{ F_n^\ell\}_{\ell=0}^\infty$ is the totality of such $F_n$.

$H$ is an infinite matrix $(H_{n\ell})$ of self-adjoint operators $H_{n\ell}$
in $\HH^n$.
Axiom 1 asserts that this matrix converges in the sense of (2).

We should repeat our standpoint here.
Our purpose in this paper is to construct at least one consistent theory
 which unites quantum mechanics and relativity.
As a mathematical problem, it is of course possible to consider the
 propagation $\exp[-iTH]$ along some global time $T$ in the grand universe
 $\UU$.
However, this does not seem to lead to any consistent unification of
quantum mechanics and relativity.
Axiom 1 is adopted so that it gives a starting point of our consistent
 unification.

\vskip 18pt

\F
{\bf 4. --
Local axioms.}
\normalsize

\vskip 8pt

What we can perceive in the world are the (local) positions and movements
 of other matters relative to our own position.
Comparing various movements, we determine time as a ratio of the positions
 and the movements (velocities).
So there exist first the local and relative positions and velocities
 or momenta.
This fact is formulated as in the following axiom.

\BP

{\it Axiom} 2.\
Let $n\ge 1$ and $F_{n+1}$ be a finite subset of
${\mbox{\bf N}}=\{1,2,\cdots\}$ with $\sharp(F_{n+1})=n+1$.
Then for any  $j\in F_{n+1}$, there exist self-adjoint
operators $X_j =(X_{j 1},X_{j2},X_{j3})$, $P_j =(P_{j1},P_{j2},P_{j3})$
 in $\HH^n$ and constants $m_j>0$ such that
\beq
[X_{j\ell},X_{k m}]=0,\q [P_{j\ell},P_{k m}]=0,\q [X_{j\ell},P_{k m}]=
i\delta_{jk}\delta_{\ell m},
\ene
\beq
\sum_{j\in F_{n+1}} m_j X_j=0,\q
\sum_{j\in F_{n+1}} P_j=0.
\ene

\noindent
By the Stone-von Neumann theorem, {\it axiom} 2 also specifies
the space dimension (see Abraham-Marsden [6], p.452).
 We identify $\HH^n$ with $\Ltnn$ in the following.

What we want to mean by $H_{n\ell}$ $(n,\ell\ge 0)$ in {\it axiom} 1
 is the $N=(n+1)$-body Hamiltonian in the usual quantum mechanics.
For the local Hamiltonian $H_{n\ell}$ we thus make the following postulate.

\BP

{\it Axiom} 3.\
Let $n\ge0$ and $F_N$ $(N=n+1)$ be a finite subset of
 ${\mbox{\bf N}}=\{1,2,\cdots\}$ with $\sharp(F_N)=N$.
Let $\{ F_N^\ell\}_{\ell=0}^\infty$ be the totality of such $F_N$.
Then the Hamiltonians $H_{n\ell}$ $(\ell\ge0)$ are of the form
\beq
H_{n\ell}=H_{n\ell0}+V_{n\ell},\q V_{n\ell}=
\sum_{{\scriptstyle \alpha=(i,j)}
\atop {\scriptstyle {1\le i<j<\infty, i,j\in F_N^\ell} } }
 V_\alpha(x_\alpha)
\ene
on $C_0^\infty(R^{3n})$, where $x_\alpha=x_i-x_j$ with $x_i$ being
the position vector of the $i$-th particle, and $V_\alpha(x_\alpha)$
is a real-valued measurable function of $x_\alpha\in R^3$ which is
$H_{n\ell0}$-bounded with $H_{n\ell0}$-bound of $V_{n\ell}$ less than 1.
$H_{n\ell0}=H_{(N-1)\ell0}$ is the free Hamiltonian of the $N$-particle
 system.
The concrete form is expressed as in (1.4) of ref. [7], if one uses
clustered Jacobi coordinates.

\BP

This axiom implies that $H_{n\ell}=H_{(N-1)\ell}$ is uniquely extended
to a self-adjoint operator in $\HH^n=\HH^{N-1} =L^2(R^{3(N-1)})$ by
the Kato-Rellich theorem.

\vskip 18pt

\F
{\bf 5. --
Local times.}
\normalsize

\vskip 8pt

For the $N$-body Hamiltonian $H_{N-1}=H_{n\ell}$ $(N=n+1)$ the following
theorem 1 is known [8] to hold under suitable assumptions on the pair
 potentials (assumption 1 in [9]) and on the decay property of subsystem
 eigenvectors (assumption 2 in [9]).

We here follow the notation and conventions in [7,9] for the $N$-body
quantum systems.
In particular $H_b=H_{(N-1)b}=H_{N-1}-I_b=H^b_{n\ell} +T_{n\ell b}=H^b+T_b$
 is the truncated Hamiltonian for the cluster decomposition
$1\le\ve b\ve\le N$, and  $P_b^M$ is the $M$-dimensional partial projection
 of the eigenprojection $P_b=P_{H^b}$ associated with the subsystem $H^b$,
 {\it i.e.} $P_b$ is the orthogonal projection in $\HH^b=L^2(R^{3(N-\vert
 b\vert )})$ onto the eigenspace of $H^b$.
$q_b$ is the velocity conjugate to the intercluster coordinates $x_b$.
We define for a $k$-dimensional multi-index   $M=(M_1,\cdots,M_k)$,
$ M_j \ge 1$,
\beq
   {\hat P} ^M_{k}= \left(I-\sum_{\vert b \vert = k}P_b^{M_k}\right) \cdots
   \left(I-\sum_{\vert d \vert = 2}
   P_d^{M_{2}}\right)
   (I-P^{M_1}),
\quad
      k=1,\cdots ,N-1,
\ene
where $P^{M_1}=P_a^{M_1}$ with $\ve a\ve=1$, and for a $\ve b\ve$-dimensional
 multi-index $M_{b} = (M_1, \cdots ,$
$ M_{\vert b \vert -1}, M_{\vert b \vert})$
$ = ({\hat M}_{b}, M_{\vert b \vert})$
\beq
   {\tilde P}_{b}^{M_{b}}=P_b^{M_{\vert b \vert}}{\hat P}_{\vert b \vert
   -1}^{{\hat M}_{b}}, \q 2\le\ve b\ve\le N.
\ene
It is clear that
\beq
\sum_{2\le\ve b\ve\le N} {\tilde P}^{M_{b}}_{b} = I - P^{M_1},
\ene
provided that the component $M_k$ of $M_{b}$ depends only on the number
 $k$ but not on $b$.
In the following we use such $M_{b}$'s only.
Under these circumstances, the following is known to hold.
\BP

{\it Theorem} 1 ([8]).
{\it Let assumptions} 1 {\it and} 2 {\it in} [9] {\it be satisfied.
Let} $f \in \HH^{N-1}$.
{\it Then there is a sequence} $t_m \to \pm\infty$ ({\it as} $m
 \to \pm\infty)$ {\it and a sequence} $M^m_{b}$ {\it of multi-indices
 whose components all tend to} $\infty$ as $m \to \pm \infty$
{\it such that for all cluster decompositions} $b$, $2\le\ve b\ve\le N$,
 $\psi \in C_0^\infty (R^1)$, {\it and} $ \varphi \in C_0^\infty
(R^{3(\vert b \vert -1)})$,
\beq
   \left\Vert \frac {\vert x^b \vert ^2}{t^2_m} {\tilde P}_{b}^{M^m_{b}}
   \exp[-it_m H_{N-1}] f \right\Vert \to 0,
\ene
\beq
\Vert \{ \psi (H_{N-1}) - \psi (H_b) \} {\tilde
   P}_{b}^{M^m_{b}} \exp[-it_m H_{N-1}] f \Vert \to 0,
\ene
\beq
   \Vert \{ \varphi (x_b/t_m) - \varphi (q_b) \} {\tilde P}_{b}^{M^m_{b}}
   \exp[-it_m H_{N-1}] f \Vert \to 0
\ene
{\it as} $m \to \pm\infty$.

\BP

{\it Definition} 1.\
Let $\phi=(\phi_{n\ell})$ with
$\phi_{n\ell}=\phi_{n\ell} (x_1,\cdots,x_n)\in \Ltnn$
be the universe in {\it axiom} 1.
We define $\HH_{n\ell}$
as the sub-Hilbert space of $\HH^n$ generated by the functions
$\phi_{mk}(x^{(\ell)},y)$ of $x^{(\ell)}\in R^{3n}$, by regarding $y\in
R^{3(m-n)}$ as a parameter, where $m\ge n$,
$F_{n+1}^\ell\subset F_{m+1}^{k}$,
 and $x^{(\ell)}$ are the (relative) coordinates of $(n+1)$ particles in
$F_{n+1}^\ell$.  $\HH_{n\ell}$ is called a local universe of  $\phi$.
$\HH_{n\ell}$ is said to be non-trivial if $(I-P_{H_{n\ell}})\HH_{n\ell}\ne
\{0\}$.

\BP

The total universe $\phi$ is a single element in $\UU$. The local universe
$\HH_{n\ell}$ may be richer.
This is because we consider the subsystems of the universe consisting
 of a finite number of particles.
These subsystems receive the influence from the other particles of
infinite number outside the subsystems, and may vary to constitute a
 non-trivial subspace $\HH_{n\ell}$.

\BP

{\it Definition} 2.\
The restriction of $H$ to $\HH_{n\ell}$ is also denoted by the same notation
  $H_{n\ell}$ as the $(n,\ell)$-th component of $H$.
We call the pair $(H_{n\ell},\HH_{n\ell})$  a local system.
The unitary group $\exp[-itH_{n\ell}]$ $(t\in R^1)$ on $\HH_{n\ell}$ is called
 the {\it proper clock} of the local system $(H_{n\ell},\HH_{n\ell})$,
 if $\HH_{n\ell}$ is non-trivial: $(I-P_{H_{n\ell}})\HH_{n\ell}\ne \{0\}$.
 (Note that the clock is defined only for $N=n+1\ge 2$, since $H_{0\ell}=0$.)
 The universe $\phi$ is called {\it rich} if $\HH_{n\ell}$ equals
$\HH^n=L^2(R^{3n})$ for all $n\ge 1$, $\ell\ge0$.
For a rich universe $\phi$, $H_{n\ell}$ equals the $(n,\ell)$-th component
of $H$.

\BP

The formula (11) indicates that $t_m$ is asymptotically equal to
$\pm\ve x_b\ve/\ve q_b\ve$ as $m\to\pm\infty$, independently of the choice
 of cluster decompositions $b$.
 This is precisely the actual procedure of measuring the time $t_m$ in
mechanics.
 The implication of this theorem is therefore interpreted as follows:
 If one $\langle\langle$measures$\rangle\rangle$ the time of a state
$f\in (I-P_{H_{(N-1)\ell}})\HH_{(N-1) \ell}-\{0\}$ in the local system
$(H_{(N-1)\ell},\HH_{(N-1)\ell})$ by the associated proper clock
$exp[-itH_{(N-1)\ell}]f$, namely if one measures the quotient
$\pm\ve x_b\ve/\ve q_b\ve$ of the scattered particles which are regarded
as moving  almost in a steady velocity, then that time is
asymptotically equal to the parameter $t_m$ in the exponent of
$\exp[-it_m H_{(N-1)\ell}]f$  as $m\to\pm\infty$. In this sense $t_m$
is interpreted as the quantum-mechanical proper time of the local system
 $(H_{n\ell},\HH_{n\ell})=(H_{(N-1)\ell}, \HH_{(N-1)\ell})$,
if $(I-P_{H_{(N-1)\ell}})\HH_{(N-1)\ell}\ne \{0\}$.

\BP

{\it Definition} 3.\
The parameter $t$ in the exponent of the proper clock
 $\exp[-itH_{n\ell}]=\exp[-itH_{(N-1)\ell}]$ of the local system
 $(H_{n\ell},\HH_{n\ell})$
is called the (quantum-mechanical) {\it proper time} or {\it local time}
 of the local system $(H_{n\ell}, \HH_{n\ell})$,
if $(I-P_{H_{n\ell}})\HH_{n\ell}\ne \{0\}$.
This time $t$ is denoted by $t_{(H_{n\ell},\HH_{n\ell})}$ indicating the
 local system under consideration.

\BP

This definition is the reverse to the usual definition of the motion or
 dynamics of the $N$-body quantum systems, where the time $t$ is given
 {\it a priori} and the motion of the particles is defined by
$\exp[-itH_{(N-1)\ell}]f$ for a given initial state $f$ of the system.

We notice here that there are two possible
$\langle\langle$directions$\rangle\rangle$ or
$\langle\langle$orientations$\rangle\rangle$ of time $t$;
the one where $t$ increases to $+\infty$, and the other where $t$
 decreases to $-\infty$.
 So far discussed, one can choose an arbitrary orientation from
 them depending on each local system.
However, the {\it axiom} 4 (the general principle of relativity)
 which will be introduced below determines the orientation of time
 to be common to {\it all}  local systems (see Hawking-Ellis [10], p.181).

{\it Time} is thus defined only for the local systems
$(H_{n\ell},\HH_{n\ell})$  and is determined by the associated proper
  clock $\exp[-itH_{n\ell}]$.
 Therefore,  there are infinitely many times
$t=t_{(H_{n\ell},\HH_{n\ell})}$ each of which is proper to
the local system $(H_{n\ell},\HH_{n\ell})$.
 In this sense time is a local notion.
 There is no time for the total universe $\phi$ in {\it axiom} 1,
 which is a (stationary) bound state for the total Hamiltonian $H$.

This local time is an approximate one in a double sense: First, $t_m$
 is only {\it asymptotically} equal to $\pm\ve x_b\ve/\ve q_b\ve$ as
 $m\to\pm\infty$.
This fact explains the so-called principle of uncertainty in our context.
In the usual explanation, the position $x_b$ and the velocity $q_b$ or
 the momentum $p_b$ cannot be determined with equal accuracy.
According to our theory, this is rephrased as follows: The time $t$ cannot
 be determined accurately, even if $x_b$ and $q_b$ could be determined
 precisely.
It is only determined in some {\it mean} sense as in (11).
(See Enss [8].
Also see Derezi\'nski [11], sect. 5, for more precise inequalities
which hold for $x_b$, $q_b$ and the local time $t$ under some
decompositions of the phase space other than (8).)
Second, the local Hamiltonian $H_{n\ell}$ is not the total Hamiltonian $H$.
Or rather, the time arises from this approximation of $H$ by $H_{n\ell}$.
This approximation may make $\HH_{n\ell}$ non-trivial, and the clock
$\exp[-itH_{n\ell}]$ can be defined as in {\it definition} 2 owing to
 $(I-P_{H_{n\ell}})\HH_{n\ell}\ne \{0\}$. On the contrary, the total
 universe $\phi$ has no associated clock and time, since $(I-P_H)\phi=0$.

Our theory of local times further implies in particular that local
systems $(H_{n\ell},\HH_{n\ell})$ cannot be decomposed into pieces and
 are {\it mutually independent}: It is true that for a subset
$F_{N^\prime}^{\ell^\prime}\subset F_N^\ell$ with $N^\prime< N$,
$H_{(N^\prime-1)\ell^\prime}$ is a subsystem Hamiltonian of $H_{(N-1)\ell}$.
However, the corresponding times $t_{N^\prime\ell^\prime}$ and $t_{N\ell}$
are measured mutually independently  as in theorem 1.
Namely, the clocks $\exp[-itH_{(N^\prime-1)\ell^\prime}]$ and
$\exp[-itH_{(N-1)\ell}]$ are different.
 More precisely speaking, the base Hilbert spaces
 $\HH_{(N^\prime-1)\ell^\prime}$ and $\HH_{(N-1)\ell}$ have different
 representations $L^2(R^{3(N^\prime-1)};N^\prime,\ell^\prime)$ and
$L^2(R^{3(N-1)};N,\ell)$ in general even on the common configuration space
 $R^{3(N^\prime-1)}_{x^{(\ell^\prime)}}$.
Thus the corresponding $\ve x^\prime_b\ve/\ve q^\prime_b\ve$ and
$\ve x_b\ve/\ve q_b\ve$ are not correlated in general.
 The same is true for two arbitrary different local systems
 $(H_{n\ell},\HH_{n\ell})$ and $(H_{mk},\HH_{mk})$.

The present theory differs from most of the existing theories of
the universe in this point.
 They start from the existence of particles or material points obtained
 by dividing matters into pieces infinitesimally.
 In our theory the local systems are regarded as the generic points or as
 the classical particles and cannot be divided further, because any
 divisions vary the associated space-time correspondingly.
 In this sense each local system $(H_{n\ell}, \HH_{n\ell})$ is a
 $\langle\langle$glocal$\rangle\rangle$ existence: It is neither
 a local thing nor a global one.

We have defined the (local) time $t=t_{(H_{n\ell},\HH_{n\ell})}$ for each
 local system $(H_{n\ell},\HH_{n\ell})$. This time $t$ satisfies theorem
 1-(11).
If one regards the time $t$ as a {\it given} quantity, this fact is
interpreted as follows: In each local system $(H_{n\ell},\HH_{n\ell})$,
 physics follows quantum mechanics, {\it i.e.} it follows the Schr\" odinger
 equation.

Our definition of times is consistent with the theory of (general)
 relativity of Einstein.
Our (quantum-mechanical) proper time of the local system
 $(H_{n\ell}, \HH_{n\ell})$ can be regarded as the quantum-mechanical
 correspondent to the classical proper time in the theory of relativity.
Within the local system  $(H_{n\ell},\HH_{n\ell})$ the velocity $q_b$
 can be arbitrarily large.
 This is  because one uses the proper time of the system
$(H_{n\ell},\HH_{n\ell})$.
If one measures the velocity of other systems from one's own system using
 the associated proper time, then those other systems move in accordance with
 the relativity theory.

\vskip 12pt

\F
{\bf 6. --
Relativity.}
\normalsize

\vskip 8pt

For the relative motions of the {\it centres} of mass of local systems, we
 postulate the principle of (general) relativity and the principle of
 equivalence as in Einstein [12].

What should be stated first on our introduction of relativity is that
 {\it only} the relative classical motions of the {\it centres} of mass of
 local systems are {\it observable} in our theory.
The {\it internal} quantum-mechanical motion within each local system is
independent of {\it observation}, at least at the present stage of our theory
till sect. 9.
In this sense, the internal quantum-mechanical motion within a local
system is {\it unobservable}.
We postulate {\it axiom} 6 in sect. 9, which gives a principle of
the {\it deduction} of the {\it internal} quantum-mechanical motion
 within each local system from classical observations of its {\it sub}local
 systems, through certain relativistic considerations.

Following {\it definition} 3 and eqs.(4) of {\it axiom} 2, we define the
local space-time
$(t,x)=(t_{(H_{n\ell},\HH_{n\ell})},$ $x_{(H_{n\ell},\HH_{n\ell})})$ with
$x=(x_1,x_2,x_3)\in R^3$ of the local system $(H_{n\ell},\HH_{n\ell})$
such that the centre of mass of the local system  $(H_{n\ell},\HH_{n\ell})$
is at
the {\it origin} of the space coordinates $x=(x_1,x_2,x_3)\in R^3$.
In this sense time $t=t_{(H_{n\ell},\HH_{n\ell})}$ is interpreted as the
 relativistic proper time associated with the centre of mass of the system
 $(H_{n\ell}, \HH_{n\ell})$.
Once the local space-time $(t,x)=(t_{(H_{n\ell},\HH_{n\ell})}, x_{(H_{n\ell},
 \HH_{n\ell})})$ has been defined, one can {\it observe} the movements of
 other systems from {\it this} space-time coordinates.
The space-time $(t_{(H_{mk}, \HH_{mk})}, x_{(H_{mk},\HH_{mk})})$ of the
 other system $(H_{mk},\HH_{mk})$ is defined {\it independently of} that of
 system $(H_{n\ell},\HH_{n\ell})$.
There is no quantum-mechanical correlation between two local-coordinate
 systems $(t_{(H_{n\ell},\HH_{n\ell})},$ $x_{(H_{n\ell},\HH_{n\ell})})$ and
 $(t_{(H_{mk},\HH_{mk})}, x_{(H_{mk},\HH_{mk})})$ unless one, or the observer,
 unites those two systems $(H_{n\ell},\HH_{n\ell})$ and $(H_{mk},\HH_{mk})$
in a single system for his particular purposes of observation.
(We remark that the combined or united system exists a priori by
{\it definition} 2, independently of the observer's concern.) In this case,
  the coordinates of the resulting combined system $(H_{pj},\HH_{pj})$ is
 again independent of those of the subsystems $(H_{n\ell},\HH_{n\ell})$ and
 $(H_{mk},\HH_{mk})$.
Thus the quantum mechanics which governs the inside of each local system
 puts no restriction on the relative motions of the centres of mass of local
 systems, as will be seen in theorem 2 below.
Therefore, there is no reason to exclude classical mechanics in describing
 the relative motions of the {\it centres} of mass of other systems
{\it observed} in one's coordinate system
$(t,x)= (t_{(H_{n\ell},\HH_{n\ell})}, x_{(H_{n\ell},\HH_{n\ell})})$.

\MP

{\it Axiom} 4.\
Those laws of physics which control the {\it relative} motions of the
{\it centres} of mass of the {\it observed} local systems are expressed by
 the classical equations which are covariant under the change of
{\it observer's} coordinate systems of $R^4$:
 $(t,x)=(t_{(H_{mk},\HH_{mk})},x_{(H_{mk},\HH_{mk})})$ to
$(t,x)=(t_{(H_{n\ell},\HH_{n\ell})},x_{(H_{n\ell},\HH_{n\ell})})$
 for any pairs $(m,k)$,
 $(n,\ell)$.

\MP

It is included in this axiom that one can observe the positions of
other systems ({\it i.e.} their centres of mass) in his coordinate system
 $(t,x)$.
 The relative velocities of the observed systems are then defined as
 quotients of the relative positions of those systems and the (local and
 quantum-mechanical) time $t$ of one's own or the observer's system.
 These are our definitions of the measurement procedure of {\it classical}
 quantities, which accord with the ordinary (implicit) agreement among
physicists where the time is given a priori.
(This is the point where we reversed the order of the time and the velocity
 in our definitions of quantum-mechanical times.)

\MP

{\it Axiom} 5.\
The coordinate system $(t_{(H_{n\ell},\HH_{n\ell})}, x_{(H_{n\ell},
 \HH_{n\ell})})$ associated with the local system  $(H_{n\ell},\HH_{n\ell})$
 is the  local Lorentz system of coordinates.
 Namely, the gravitational potentials $g_{\mu\nu}$ for the {\it centre} of
 mass of the local system  $(H_{n\ell}, \HH_{n\ell})$, {\it observed} in
these coordinates
$(t_{(H_{n\ell}, \HH_{n\ell})}, x_{(H_{n\ell},\HH_{n\ell})})$, are equal
 to $\eta_{\mu\nu}$.
Here $\eta_{\mu\nu}=0$ $(\mu\ne\nu)$, $=1$ $(\mu=\nu=1,2,3)$, and $=-1$
$(\mu=\nu=0)$.

\MP

For the field equation which determines the metric $g_{\mu\nu}$, we refer to
 Hawking-Ellis [10], p.74 or Friedman [3], p.180.

{\it Axiom} 5 implies that for the coordinate system
 $(t_{(H_{n\ell},\HH_{n\ell})}, x_{(H_{n\ell},\HH_{n\ell})})$ associated
 with the local system  $(H_{n\ell}, \HH_{n\ell})$,  the principle of
 constancy of light velocity holds in the following special sense:
 The light radiated from another system $(H_{mk}, \HH_{mk})$ moving with
 a steady velocity relative to one's own system $(H_{n\ell},\HH_{n\ell})$
 propagates through the {\it flat} region where $g_{\mu\nu}=\eta_{\mu\nu}$,
 at a constant speed independent of the velocity of the system
$(H_{mk}, \HH_{mk})$ relative to  one's own system
$(H_{n\ell}, \HH_{n\ell})$.

\BP

{\it Theorem} 2. \
{\it Axioms} 4 {\it and} 5 {\it are consistent with axioms} 1-3.

\MP

{\it Proof}.
This may be clear by the mutual independence of local systems and
the associated times stated at the end of the previous section: We can
assume any equations on relative motions, which follow arbitrarily given
 transformation rules under the change of the coordinates systems between
 any two local systems, and give arbitrary numbers to $g_{\mu\nu}$, as far
 as these are consistent with
$\langle\langle$physics$\rangle\rangle$.
 Whichever equations may one adopt to govern the relative motions of
centres of mass of local systems, these equations are consistent with
the quantum mechanics that controls the inside of the local systems.

For the sake of clarity, however, we repeat the argument of the previous
 section.
The local coordinate system
$(t_{(H_{n\ell},\HH_{n\ell})}, x_{(H_{n\ell},\HH_{n\ell})})$
 is determined only within each local system $(H_{n\ell},\HH_{n\ell})$,
 through the quantum-mechanical {\it internal} motions of the system.
This coordinate system is independent of the local coordinate system
 $(t_{(H_{mk},\HH_{mk})},x_{(H_{mk},\HH_{mk})})$  of any other local system
 $(H_{mk},\HH_{mk})$.
 This is due to the mutual  independence of the $L^2$ representations of
 the base Hilbert spaces $\HH_{n\ell}$ and $\HH_{mk}$.

The relativity axioms, {\it axioms} 4 and 5, are concerned merely with the
{\it centres} of mass of local systems $(H_{mk},\HH_{mk})$, {\it observed}
by an observer system $(H_{n\ell},\HH_{n\ell})$ with coordinate system
 $(t_{(H_{n\ell},\HH_{n\ell})}, x_{(H_{n\ell},\HH_{n\ell})})$.
 This {\it observer's} coordinate system $(t_{(H_{n\ell},\HH_{n\ell})},
 x_{(H_{n\ell},\HH_{n\ell})})$ is {\it independent of} the coordinate system
 $(t_{(H_{mk},\HH_{mk})}, x_{(H_{mk},\HH_{mk})})$ of the {\it observed}
 system $(H_{mk},\HH_{mk})$, as stated in the previous paragraph.
Because of this independence, the system $(H_{mk},\HH_{mk})$ can follow
 quantum mechanics ({\it axioms} 1-3) {\it inside} the system
{\it with respect to} its own coordinate system $(t_{(H_{mk},\HH_{mk})},
 x_{(H_{mk},\HH_{mk})})$, {\it as well as} its {\it centre} of mass
 can follow     general relativity ({\it axiom}  4) or any other given
 postulates {\it with respect to} the observer's coordinate system
 $(t_{(H_{n\ell},\HH_{n\ell})}, x_{(H_{n\ell},\HH_{n\ell})})$.
 This is the case, even if the coordinate system
 $(t_{(H_{n\ell},\HH_{n\ell})},  x_{(H_{n\ell},\HH_{n\ell})})$ of the
 observer coincides with the coordinate system $(t_{(H_{mk},\HH_{mk})},
 x_{(H_{mk},\HH_{mk})})$ of the observed system itself, because the motion
 of the {\it centre} of mass and the internal {\it relative} motion of a local
 system are mutually independent.
Therefore, the local Lorentz postulate ({\it axiom} 5) of the {\it centre}
of mass of the system $(H_{n\ell},\HH_{n\ell})$ also does not contradict the
 Euclidean postulates in sect. 3-5 of the internal space-time of that system.

In this sense, {\it axioms} 4 and 5 are chosen so that the relativity theory
 holds between the {\it observed} motions of {\it centres} of mass of local
systems, and {\it have nothing to do with} the {\it internal} motion of each
 local system, which obeys {\it axioms} 1-3.
Thus {\it axioms} 4 and 5 are consistent with {\it axioms} 1-3.   $\Box$

\SP

{\it Remark.}
As stated in the introduction, the theory of local times can be reformulated
 or  redescribed as a kind of vector bundle theory with ten-dimensional total
 space $X \times R^6$ and four-dimensional base space $X$.
 $R^6$ corresponds to the Euclidean space of quantum mechanics inside the
 local systems.
$X$ is the Riemannian manifold with metric $g_{\mu\nu}$, the gravitational
 potentials.
$X$ corresponds to the classical space-time observed by a fixed observer,
 which gives the observer's reference frame or coordinate  system for
 measurements of the classical particles, the centres of mass of other local
 systems.
Theorem 1 due to Enss is interpreted as a  contraction procedure of the fibre
 $R^6$ to four-dimensional space-time $R^4$ in an approximate context of the
 uncertainty principle described in sect. 5 after {\it definition} 3.
Owing to the orthogonality of $X$ and $R^6$ in the total space $X \times R^6$,
 the quantum and relativity theories hold in $R^6$ and in $X$,
 respectively, without mutual contradiction.

\SP

This is the mathematical explanation of the consistency of {\it axioms}
 1-5, namely the proof of theorem 2.

This proof describes the way of the
 $\langle\langle$orthogonalization$\rangle\rangle$ of
 quantum mechanics and relativity, as announced in sect. 2.

Under {\it axiom} 5, the local time $t_{(H_{n\ell},\HH_{n\ell})}$ of
 the system $(H_{n\ell},\HH_{n\ell})$ coincides with the relativistic
proper time of the centre of mass of the system, because, at the centre
 of mass, the space coordinates $x_{(H_{n\ell},\HH_{n\ell})}=0$.

The relation of the {\it internal} motion of a local system with the motion
 of {\it centres} of mass of its {\it sub}systems will be discussed in sect.
 9 which deals with observation.

Summing up, we have obtained the following physical picture of the universe:
 The universe $\phi$ is a stationary bound state of the total Hamiltonian
 $H$ with infinite degrees of freedom.
Times $t=t_{(H_{n\ell},\HH_{n\ell})}$ appear only for local systems
 $(H_{n\ell},\HH_{n\ell})$.
 According to this time $t$, the physics laws within the local system
 $(H_{n\ell},\HH_{n\ell})$ obey the quantum mechanics, and the physics laws
 outside the system which govern the relative motions between the centres of
 mass of local systems obey the classical theory of relativity.
 These two sorts of view are consistent, because the quantum mechanics
 inside the local systems puts no restriction on the motions of the centres
 of mass of local systems owing to the mutual independence of local systems.

\vskip 18pt

\F
{\bf 7. --
A paradox of cyclotron.}
\normalsize

\vskip 8pt

As an illustration of the unification of quantum mechanics and relativity
in our context, we consider the experiment by a cyclotron.
To clarify the point of our argument, we consider an ideal situation.

We suppose that some electrons for example are accelerated by a cyclotron,
 and can have velocities very near the velocity of light.
 By this experiment the observer can $\langle\langle$see$\rangle\rangle$
 many phenomena: the electrons are accelerated to have the velocities near
the velocity of light, hence some relativistic phenomena occur, and some of
 the electrons may hit a nucleus and they together produce or change into
 several elementary particles.
{}From these observations, the observer $\langle\langle$knows$\rangle\rangle$
 the masses, velocities, energies, and so on, of these particles by some
 reasoning or analysis of the experimental data.
In this experiment, the quantum-mechanical and relativistic effects look
like appearing at the same $\langle\langle$time$\rangle\rangle$.
This might be taken as a contradiction: As we shall see in the next section,
quantum mechanics yields     non-locality, which contradicts the local
 causality deduced from relativity.

The answer to this problem is as follows:
The local causality is only concerned with the {\it observed} relative
 motions between the centres of mass of local systems, {\it e.g.}, between
 the system $(H_{n\ell},\HH_{n\ell})$ of the electrons under the
acceleration, and the system $(H_{mk},\HH_{mk})$ of the stationary nucleus.
 In this case, the {\it observed} motion between the two systems naturally
follows the relativity theory, and the causality holds, according to our
 theory.
When the observer $\langle\langle$watches$\rangle\rangle$ the collision of
 the accelerated electrons and the stationary nucleus, it looks at the inside
 of the combined local system $(H_{pj},\HH_{pj})$ of the two systems
$(H_{n\ell},\HH_{n\ell})$  and $(H_{mk},\HH_{mk})$. Since what can be observed
 is only the relative classical motions of the centres of mass of sublocal
  systems of $(H_{pj},\HH_{pj})$,
 the observer cannot see the inside of the system $(H_{pj},\HH_{pj})$.
However, as we shall discuss in sect. 9,   one can deduce the internal
 quantum-mechanical motion within $(H_{pj},\HH_{pj})$ from these
 observations of sublocal systems.
This deduction leads one to conclude that the physics {\it within} the system
$(H_{pj},\HH_{pj})$ follows quantum mechanics, and he sees the
quantum-mechanical phenomena of collision in $(H_{pj},\HH_{pj})$.

These two kinds of $\langle\langle$observation$\rangle\rangle$ are
 mutually consistent in the sense described in the proof of theorem 2.
The {\it observed} motion {\it between} the centres of mass of the two
systems $(H_{n\ell},\HH_{n\ell})$ and $(H_{mk},\HH_{mk})$ can follow
classical relativity and causality, without any contradiction with the
 non-locality of quantum mechanics {\it within} the combined system
$(H_{pj},\HH_{pj})$.
  This consistency which follows from the mutual independence of the
 coordinate systems of the observer and the observed local systems is our
solution of the unification of quantum mechanics and relativity.
We shall analyse the problem in more detail in the following sections.

\vskip 18pt

\F
{\bf 8. --}  EPR {\bf paradox.}
\normalsize

\vskip 8pt

Einstein-Podolsky-Rosen [14] argued that
quantum mechanics (QM) combined with the locality principle
 (Redhead [4], p.75):

\BP

\ \ \ L) \
 Elements of reality pertaining to one system cannot be affected by
 measurements performed $\langle\langle$at a distance$\rangle\rangle$ on
 another system,

\BP

\F
implies the incompleteness of quantum mechanics.
Here completeness means that ([14])

\BP

\ \ \ C) \
Every element of the physical reality must have a counterpart in the
 physical theory.

\BP

\F
This argument is rephrased as follows:
\beq
{\rm {QM}}\Rightarrow \lnot {\rm {(L)}}\ \  \ {\rm {or}}\ \ \  \lnot
 {\rm {(C)}},
\ene
which is called {\it Einstein Dilemma} in Redhead [4].
Here QM is the theory of quantum mechanics, and
$\langle\langle\lnot\rangle\rangle$ denotes the negation.
We refer to Jammer [15], Selleri [16], Schommers [17] for further references.

Contrary to Einstein-Podolsky-Rosen [14], we adopt the standpoint that
 quantum mechanics is complete in the sense of C).
Thus quantum mechanics yields that non-locality holds within each local
 system.
(There are arguments that the negation of the Bell inequality implies
 non-locality.
Thus, in these arguments, quantum mechanics automatically yields
 non-locality.
See d'Espagnat's paper in Schommers [17].
However, there are refutations to this type of arguments.
See Redhead [4], chapt. 4 for instance.)
 In this sense, quantum mechanics contradicts
local causality in the relativity theory (see the arguments in Redhead [4],
 p.75, also see Hawking-Ellis [10], chapt. 3).
However, this situation is not a contradiction in our context of local times.
The reason is that, in our theory, the classical theory of relativity is
 concerned {\it only} with the relative motions {\it between
 the centres} of mass of local systems, {\it observed} in the
 {\it observer's} coordinate system $(t,x)$.
Thus,     local causality is required {\it only} to the {\it observed}
 relative motions of the {\it centres} of mass of local systems,
 {\it not} to the physics {\it within} the local systems, which is
{\it unobservable} directly according to our fundamental assumption made
 in sect. 2.
 This allows us to admit the {\it unobservable} non-locality {\it within}
 each local system, where quantum mechanics holds consistently with
 non-locality (by (12)).

For illustration, let us take the well-known example of two photons
 polarized mutually orthogonally.
(See Redhead [4], chapt. 3 and sect. 4.5.)
Let $(H_{n\ell},\HH_{n\ell})$ and $(H_{mk},\HH_{mk})$ be the local
 systems of each of these two photons, and let $(H_{pj},\HH_{pj})$ be
 the combined system of the  two photons.
 According to our theory, within this local system
$(H_{pj}, \HH_{pj})$ of two photons, the interactions propagate with
{\it infinite} speed by the quantum mechanics valid inside the system,
 as far as the two photons are considered to constitute one local system
 $(H_{pj},\HH_{pj})$, and are $\langle\langle$observed$\rangle\rangle$
from their creation to the measurements of their polarization traced
along the proper local time $t_{(H_{pj},\HH_{pj})}$.
In this sense, non-locality is no contradiction within the combined local
 system of the two photons.
Only when the two photons are considered as two different local systems
  $(H_{n\ell},\HH_{n\ell})$ and $(H_{mk},\HH_{mk})$ and are {\it observed}
 in the observer's coordinate system, the classical-mechanical speed of
 interactions appears and the situation looks like a contradiction.
However, since $(H_{pj},\HH_{pj})$, $(H_{n\ell},\HH_{n\ell})$,
  and $(H_{mk},\HH_{mk})$ are mutually independent quantum-mechanically
 in their respective local times, this situation is not a contradiction:
 The postulate that the relative motions between the centres of mass of
 $(H_{n\ell}, \HH_{n\ell})$ and $(H_{mk},\HH_{mk})$ observed in the
 observer's coordinates is governed by classical relativity, hence the
 local causality holds between the two systems, {\it does not contradict},
 by the mutual independence of the coordinates of these two systems,
 the fact that the combined system
$(H_{pj}, \HH_{pj})$ follows
   quantum mechanics in its own local time $t_{(H_{pj}, \HH_{pj})}$,
according to which non-locality holds.

We remark that, in this solution of the EPR paradox, no
$\langle\langle$hidden variables$\rangle\rangle$ are used.

\vskip 18pt

\F
{\bf 9. --
Observation.}
\normalsize

\vskip 8pt

So far we have postulated five axioms, which describe the fundamental
 properties of the universe, and we have seen that these axioms are
 mutually consistent.
We turn to the considerations of the observation process by an observer's
 system.

When the observer sees the universe, the potentials $V_\alpha(x_\alpha)$ of
 the {\it observed} systems effective {\it between} these local systems do
 not give any {\it quantum-mechanical} affection but should be interpreted
 as the classical potentials:
 A local system $(H_{n\ell},\HH_{n\ell})$ is closed when it is considered
 within its own proper coordinate system $(t_{(H_{n\ell}, \HH_{n\ell})},$
 $ x_{(H_{n\ell},\HH_{n\ell})})$, since the time
 $t_{(H_{n\ell},\HH_{n\ell})}$ is determined only within the system by
the associated clock $\exp[-itH_{n\ell}]$.
Other local systems $(H_{mk},\HH_{mk})$ can influence the
local system $(H_{n\ell},\HH_{n\ell})$ in the quantum-mechanical
 way only when these are considered by an observer as a single combined
 system of $(H_{n\ell},\HH_{n\ell})$ and $(H_{mk},\HH_{mk})$.
 Only at this stage the time is common to these two systems, and the
 quantum-mechanical correlation between these systems through pair
 potentials $V_\alpha(x_\alpha)$ can be discussed on the same scales of
 space-time.
The principle of uncertainty in the sense stated before within the united
 system of $(H_{n\ell},\HH_{n\ell})$ and $(H_{mk},\HH_{mk})$ holds only
 at this stage.
{\it Between} the two independent systems $(H_{n\ell},\HH_{n\ell})$ and
 $(H_{mk},\HH_{mk})$, the physics seems to be following
classical mechanics for the observer, and the pair potentials
$V_\alpha(x_\alpha)$ effective between these two systems operate as
the classical ones as in electromagnetism.
One system $(H_{n\ell},\HH_{n\ell})$ of these systems plays the role of
 the observer system, and another $(H_{mk},\HH_{mk})$ is an observed system.
In this case the coordinate system is
 $(x^\lambda)_{\lambda=0}^3=(t,x) =(t_{(H_{n\ell},\HH_{n\ell})},
 x_{(H_{n\ell},\HH_{n\ell})})$.
By {\it axioms} 4 and 5 the motion of the centre of mass of the observed
 system relative to the observer system, when we neglect the effects other
 than the gravitational force, is thus expressed as in the theory of general
 relativity by the equation of motion
\beq
\frac{d^2 x^\lambda}{d \tau^2}+
\sum_{\mu,\nu=0}^3
\Gamma^\lambda_{\ \mu\nu}
\frac{dx^\mu}{d\tau}
\frac{dx^\nu}{d\tau}=0,
\ene
where $\tau=t_{mk}=t_{(H_{mk},\HH_{mk})}$ is the proper time for the centre
 of mass of the system $(H_{mk},\HH_{mk})$ and $\Gamma^\lambda_{\ \mu\nu}
={1}/{2}\sum_{\alpha=0}^3 g^{\lambda\alpha}
\left({\partial g_{\alpha\mu}}/{\partial x^\nu}+
 {\partial g_{\nu\alpha}}/{\partial x^\mu}
-{\partial g_{\mu\nu}}/{\partial x^\alpha}\right)$
 is the Christoffel's symbol.

We have to note that the above process of
$\langle\langle$observations$\rangle\rangle$ is concerned only with
the classical-mechanical phenomena between the centres of mass of
local systems.
As we have stated, the {\it observation} is restricted only to the classical
 relative motions of centres of mass of local systems.
However, the observer's system can also $\langle\langle$see$\rangle\rangle$
 the {\it unobservable} quantum-mechanical effects inside the observed
 system $(H_{mk},\HH_{mk})$.
Reversely to the classical-mechanical observations, the observer sees that
 the physics {\it within} the observed system is described by the
Hamiltonian $H_{mk}$ in a quantum-mechanical way.
The observer can $\langle\langle$see$\rangle\rangle$ the clock and motions
 in the observed system through the classical observations using, {\it e.g.},
 light as in the astronomical observation of electromagnetic radiations from
 stars.
The observer deduces from these classical observations, with some
 relativistic corrections of the observed values as will be discussed below,
 that the quantum-mechanical laws hold within the system $(H_{mk},\HH_{mk})$,
 {\it as far as} the time of the system is defined as the quantum-mechanical
 proper one of the system $(H_{mk},\HH_{mk})$ as in {\it definition} 3.

The results obtained in the usual relativistic quantum theories could be
 explained in our context as follows:
The quantum phenomena occurring in a local system follow
non-relativistic quantum mechanics, but the observed values of
quantum-mechanical quantities should be corrected according to the
classical relativity so that the corrected values equal the values
predicted by the (non-relativistic) quantum-mechanics.
The relativistic effects arise only related with the {\it observation}
of the classical quantities of the local system.
The so-called {\it propagation} of fields or forces is interpreted in
our theory also as the classical phenomena which appear through the
observation process.
What the observer measures concerning the observed system
$(H_{mk},\HH_{mk})$ is the classical-physics values of the subsystems
or the portions of $(H_{mk}, \HH_{mk})$, {\it e.g.}, the positions,
momenta, etc., of the subsystem's centres of mass, measured according
 to the observer's local time.
{}From these classical quantities, the observer deduces through the classical
 relativistic corrections  that     quantum mechanics is working in the
 observed system $(H_{mk}, \HH_{mk})$.

If we see this procedure in reverse order, as in the last but two
paragraph of sect. 5, beginning with quantum mechanics, {\it i.e.}
 with the Schr\"odinger propagator $\exp[-it_{mk}H_{mk}]$, given
 the system $(H_{mk}, \HH_{mk})$ with coordinates
$(t_{mk},x_{mk})=(t_{(H_{mk},\HH_{mk})},$ $x_{(H_{mk},\HH_{mk})})$,
it is described more precisely as follows:
The quantum-mechanical velocities of the particles in the system
 $(H_{mk}, \HH_{mk})$ are given by the quotients $q_b= x_b/t_{mk}$,
 asymptotically as $t_{mk}\to\infty$,  of the position vectors $x_b$
of the particles and the local time $t_{mk}$.
 This is the case, since we have assumed that the coordinates
$(t_{mk},x_{mk})$, in particular  the local time $t_{mk}$, are given
 through the propagator or clock $\exp[-it_{mk}H_{mk}]$ and {\it axiom} 2.
Our fundamental assumption here on the observation is as follows:

\BP

{\it Axiom} 6.
The momenta $p_j=m_j x_j/t_{mk}$ of the particles $j$ with mass $m_j$
 in the observed local system $(H_{mk},\HH_{mk})$ with coordinate system
 $(t_{mk},x_{mk})$, given as above, are observed, by the observer system
 $(H_{n\ell},\HH_{n\ell})$ with coordinate system $(t_{n\ell},x_{n\ell})$,
 as $p_j^\prime=m_j x_j^\prime/t_{n\ell}$, where $x_j^\prime$ is obtained
from $x_j$ by the relativistic transformation of coordinates:
  $(t_{mk},x_{mk})$ to $(t_{n\ell},x_{n\ell})$ as
in {\it axiom} 4.
The same is true for the observation of the energies of the particles:
 the energies of the particles in the observed local system are observed
by the observer as the ones transformed in accordance with the relativity.

\BP

Namely, it is assumed that the quantum-mechanical momenta
$p_j=m_j x_j/t_{mk}$ of the particles within the system $(H_{mk},\HH_{mk})$
 are observed in actual experiments by the observer system
$(H_{n\ell},\HH_{n\ell})$ with coordinate system $(t_{n\ell},x_{n\ell})$,
 as the {\it classical} quantities $p_j^\prime =m_j x_j^\prime/t_{n\ell}$
 whose values are calculated or predicted by correcting the
quantum-mechanical values $p_j$ with taking the relativistic effects
 of observation into account.
A similar assumption is made for the energies of the particles.

In this sense, {\it axiom} 6 should be called a principle of deduction
of the experimentally observed values from the non-relativistic quantum
 mechanics through certain relativistic corrections, rather than be called
 a deduction rule of the internal quantum-mechanical motions from the
 classical observations of the sublocal systems.
We adopted the latter expression in sect. 2 and 6 for the sake of simplicity
 of expression.

We should note that {\it axiom} 6 is consistent with our {\it axioms} 1-5
 in the following sense: {\it axiom} 6 is concerned only with the quantum
 mechanics {\it within} the local system $(H_{mk},\HH_{mk})$, so that it
 gives the rules to transform the quantum-mechanical values, {\it e.g.}
 $p_j$, of the system $(H_{mk},\HH_{mk})$ to the values, {\it e.g.}
 $p_j^\prime$, {\it observed} experimentally by the observer.
It is therefore not related with {\it any} physics laws of the particles
 {\it within} the system $(H_{mk},\HH_{mk})$, unless the transformed
values ({\it e.g.}  $p'_j$) are compared with the actual experimental values.
In this sense, {\it axiom} 6 is concerned only with {\it how the nature looks
 at the observer}.
Together with {\it axioms} 1-5, it gives the prediction of the physical
values observed in actual experiments, and is checked solely through the
 experimental data.

In this sense, {\it axiom} 6 together with {\it axioms} 1-5 gives our
 prediction for the relativistic observations of the quantum-mechanical
 local systems.
For the illustration, let us take an example of the calculation of the
 differential cross-section ${d\sigma}/{d\Omega}$ for the scattering
 phenomenon of an electron by a Coulomb potential ${Ze^2}/{r}$,
where $r=\vert x\vert$ and $x$ is the position vector of the electron
 relative to the scatterer.
We assume that the scatterer has a very large mass compared to the
electron and that ${Z}/{137}$ is small.
 Then, as usual, quantum mechanics gives, in a Born approximation,
\beq
\frac{d\sigma}{d\Omega}=\frac{Z^2e^4}{16E^2\sin^4(\theta/2)},
\ene
where $\theta$ is the scattering angle, and $E$ is the total energy
 of the system consisting of the electron and the scatterer.
 Thus, since the electron is far away from the scatterer after the
 scattering, we may assume that the energy $E$ is  equal to the kinetic
 energy of the electron and the scatterer.
We assume that the observer is stationary relative to this local system of
 the electron and the scatterer.
This means that the observer is approximately stationary relative to the
 scatterer, since we assumed the mass of the scatterer much larger than
 that of the electron.
{}From the observer, which is stationary relative to the scatterer, the energy
 $E$ is equal to the kinetic energy of the electron and is observed as a
 classical quantity by {\it axiom} 6.
Therefore, its actual observed value equals  $p^0-p^0_0$ by
special relativity.
Here  $p^0_0$ is the rest energy of the electron, and $p^0$ is the
relativistic energy of the electron given as follows.
Let $v$ be the absolute value of the coordinate velocity of the electron:
 $v=\vert dx/dt\vert$, where $t$ is the observer's local time.
Then, by the theory of special relativity, $p^0$ is expressed as follows:
\beq
p^0=\frac{1}{\sqrt{1-v^2}}\  p^0_0.
\ene
Here we adopted a unit system such that the speed of light $c=1(>v\ge 0)$.
Thus $p^0_0$ equals the rest mass $m_0$ of the electron: $p^0_0=m_0$.
We can then compute
\beq
E=p^0-p^0_0=\frac{1-\sqrt{1-v^2}}{\sqrt{1-v^2}}\ p^0_0
\approx \frac{m_0 v^2/2}{\sqrt{1-v^2}}
\approx \frac{m_0 v^2}{2}(1+v^2/2)
\approx \frac{m_0 v^2}{2}.
\ene
\vskip4pt

\noindent
Thus, taking the relativistic effects of the observation into account,
 we have the differential cross-section given by
\beq
\frac{d\sigma}{d\Omega}\approx\frac{Z^2e^4}{4m_0^2v^4\sin^4(\theta/2)}
(1-v^2).
\ene
\vskip4pt

\noindent
This coincides with the usual relativistic prediction of the Klein-Gordon
 equation obtained by a Born approximation, if the observer is assumed to be
 stationary relative to the scatterer.

The effect of the spin of the electron can also be included by introducing
 the spin-orbit interactions (see Mott-Massey [18], chapt. X).
In this treatment, the wave function of the electron is regarded as a
 two-dimensional vector-valued function as usual.
The result corresponding to (14) is the same as the one obtained through
 Dirac theory as far as the spin correction is concerned (Mott-Massey [18],
 chapt. X, sect. 3):
\beq
\frac{d\sigma}{d\Omega}
=\frac{Z^2e^4}{16E^2\sin^4(\theta/2)}
\left(1-\frac{2E}{m_0}\sin^2(\theta/2)\right).
\ene
\vskip4pt

\noindent
Inserting the relativistic correction (16), we get
\beq
\frac{d\sigma}{d\Omega}\approx\frac{Z^2e^4}{4m_0^2v^4\sin^4(\theta/2)}
(1-v^2\sin^2(\theta/2))(1-v^2).
\ene
\vskip4pt

\noindent
This is exactly the relativistic Dirac prediction in a Born approximation.
We can proceed to finer Born approximations as well, as in Mott-Massey [18],
 chapt. IX, sect. 4.5, and can recover the relativistic prediction of Dirac
 equation.

In sum, quantum mechanics has the intrinsic nature, and relativity is
 concerned with how     nature looks at the observer.
The observed relativistic quantum phenomena are explained as the
consequences of the relativistic effects of the observation of the
non-relativistic quantum systems.

\vskip 18pt

\F
{\bf 10. --
Concluding discussions.}
\normalsize

\vskip 8pt

The times are defined only for local systems $(H_{n\ell},\HH_{n\ell})$.
The total universe $\phi$ has no time associated.
The local times arise through the affections from other particles outside
 the local systems ({\it definitions} 1-3).
The uncertainty principle holds only within these local systems as the
 uncertainty of the local times.
    Quantum mechanics is confined within each local system  in this sense.
The quantum-mechanical phenomena between two local systems appear only
when they are combined as a single local system.
In the local system, the interaction and forces propagate with infinite
velocity or, in other words, they are {\it unobservable}.

Each local system can be the observer of other systems.
In this situation, the local systems are mutually independent in the
 sense that the associated quantum-mechanical local times are not correlated
 in general.
Therefore, there are no reasons to exclude classical mechanics in
describing the {\it observable} relative behaviour of the observed systems
 with respect to the observer.
Thus, the gravitational potentials can be introduced in accordance with
the theory of general relativity.
These potentials determine the global space-time structure around the
observer system.
Inside the observer system the space-time is Euclidean.
The observer itself cannot detect the gravitational correlation or the
space-time structure inside its own system, and the interactions and forces
 cannot be detected inside it.
On the contrary, between the local systems, the observer can detect only
 the classical-mechanical effects.
Namely, the gravitational forces appear and the quantum-mechanical
 potentials $V_\alpha(x_\alpha)$ effective outside the local systems
 operate as the classical ones between the local systems.
Nevertheless, through the media ({\it e.g.}, light in classical sense)
 which connect the observer and the observed systems and obey the
classical physics, the observer sees, through some relativistic corrections
 of the observed classical values, that the physics laws inside the other
 local systems follow quantum mechanics.

These facts are all the consequences of the introduction of
{\it local times} which are proper to each local system.
The time is neither a given thing nor a common one to the total universe.
On the contrary, there can be defined no global time.
More strongly the total universe is a (stationary) bound state of the
 total Hamiltonian $H$ of infinite degrees of freedom.
The times arise only when the observers restrict their attention to
 its subsystems as approximations of the total Hamiltonian $H$.
The universe itself is correlated within it as a bound state of $H$.
The observer always separates a subsystem from it, so to speak,
 artificially, and the (steady) motion and time appear.
Inside the subsystem this local time explains the quantum effects,
and outside the subsystem it explains the gravitation and the classical
 mechanics.
The relativistic quantum phenomena are explained as the relativistic
effects of the observation of the non-relativistic quantum systems.
All these physical phenomena occur by this artificial separation of
the universe.
The universe itself does not $\langle\langle$change$\rangle\rangle$:
 It is a stationary bound state.

\vskip 18pt

\F
{\bf 11. --
Some conjectures.}
\normalsize

\vskip 8pt

 As a conclusion, we state some conjectures which would likely hold
in our theory.

\MP

\ \ \  1) The universe $\phi$ would be confined in a local region of the
 infinite-dimensional configuration space $R^\infty$ in {\it some} sense
 as an eigenvector of the total Hamiltonian $H$:
 This is an analogy with the corresponding fact for the finite-dimensional
 Hamiltonian $H_{n\ell}$.
 The eigenvectors for this Hamiltonian are local in the sense that they
are in $L^2(R^{3n}_{x^{(\ell)}})$.
Or, more strongly, it is known that they decay exponentially with respect
to $\vert x^{(\ell)}\vert$ if the associated eigenvalues are not the
thresholds of $H_{n\ell}$.
This conjecture means that the universe is a
$\langle\langle$closed$\rangle\rangle$ or
$\langle\langle$finite$\rangle\rangle$ one in quantum-mechanical sense.
Therefore, this conjecture will eliminate the arguments on the so-called
 $\langle\langle$dark matter$\rangle\rangle$, which is supposed to exist
in the usual $\langle\langle$big-bang$\rangle\rangle$ theory to make the
 universe almost stable, {\it i.e.}  to make it have the density very close
 to the $\langle\langle$critical density$\rangle\rangle$, therefore to let
 the universe have matters enough
for stars and planets to exist.
A similar explanation will work for explaining the stability of such local
 systems $(H_{n\ell},\HH_{n\ell})$ as galaxies, clusters of galaxies, etc.,
 if one regards their actual quantum-mechanical states
$\phi_{n\ell}\in \HH_{n\ell}$ as (approximate) bound states of the local
 Hamiltonian $H_{n\ell}$, namely as
$\langle\langle$resonances$\rangle\rangle$ of $H_{n\ell}$
in some sense (cf. Jensen-Kato [19]).
 Then, the quantum-mechanical effect would explain their stability or
 $\langle\langle$locality$\rangle\rangle$ as in the explanation of the
 locality of the total universe without introducing the dark matter.

\MP

\ \ \ 2) It would be possible to accommodate bosons, fermions, photons
 ({\it i.e.}  the quantum theory of radiation), and so on in our theory
 by using sub-Fock spaces ${\cal F}=\bigoplus_{n=0}^\infty \HH^n$ of \
 $\UU=\bigoplus_{n=0}^\infty\left(\sum_{\ell=0}^\infty \HH^n\right)$.
As usual the spin of the particles can be introduced by taking the
 vector-valued representations $L^2(R^{3n}; {\mbox{\bf C}}^k)$ $(k\ge 1)$
 of $\HH^n$ in {\it axiom} 2.
($\mbox{\bf C}$ is the set of complex numbers.)

For the photons without interactions with matter, the Hamiltonians
$H_{n\ell}$ in the sense of {\it axiom} 3 should be taken as a sum of
 those for harmonic oscillators.
Since the corresponding total Hamiltonian
$H_p=\bigoplus_{n=0}^\infty H_{n\ell}$ has a complete system of
eigenvectors, one has $(I-P_{H_p}){\cal F}=\{0\}$.
 Thus it has no associated proper clock.
In this sense the light as a {\it wave} propagates with infinite speed
within the system $(H_p,\cal F)$.
It should be remarked that this statement has a different meaning than
 the statement that the light, as an electromagnetic interaction,
 propagates with infinite speed.
As such a field, light interacts with other matters through vector
potentials $A(x)$, which are quantum-mechanical electromagnetic fields
 interacting instantaneously with matters.
On the contrary, as a {\it wave}, light propagates with the constant
 speed $c$ within general local systems.
The reason is that the Hamiltonian of a general local system contains
 terms comprising the interactions between photons and other particles.
 As a consequence, that Hamiltonian does not have a complete system of
 eigenvectors, and the local time of the system is defined by
 {\it definition} 3.
Then, with some additional arguments to chapt. III of von Neumann [20],
 one can show that the speed of the wave front of light within that local
 system is the constant $c$.

In the line of this treatment of the interactions between electromagnetic
 field and matters, it would be possible to explain the Lamb shift.
 It is probable that the explanation could be given even without appealing
 to the path integral techniques described, {\it e.g.}, in Feynman-Hibbs [21].
For this possibility, we refer to the argument around the footnote 149)
 of [20], chapt. III, sect. 6, remarking that the argument there and the
one in [21], chapt. 9, are essentially the same.
As in [20], let $H$ be the total Hamiltonian of the system consisting of
 matter and radiation field ({\it i.e.}  photons), let $I$ be the
interaction between matter and photons, and let a sequence of complex
numbers $a_{{k}{M}_1{M}_2\cdots}$ belong to the space of sequences
$a_{{k}{M}_1{M}_2\cdots}$ with normalization condition that
$\sum_{k,M_1,M_2,\cdots}|a_{kM_1M_2\cdots}|^2=1$ and let
its component $a_{{k}{M}_1{M}_2\cdots}$ express the state where matter
is in the state $k$ and the number of photons in state $j$ is ${M}_j$.
 In other words, $a_{{k}{M}_1{M}_2\cdots}$ denotes the eigenvector
of $H-I$ with eigenvalue $W_k+\sum_{j=1}^\infty h\rho_j\cdot M_j$,
 where $W_k$ is the eigenenergy of the matter state $k$, $\rho_j$ is
the eigenfrequency of photons in state $j$, and $h$ is the Planck constant.
Set $b_{{k}{M}_1{M}_2\cdots}(t) =
 \exp{\{-itH\}}\exp{\{it(H-I)\}}a_{{k}{M}_1{M}_2\cdots}$.
Consider the transition process where the matter part of the total
system begins and ends in the same state ${\bar k}$.
Integrating the differential equation
\beq
\nonumber
\frac{1}{i}\frac{d}{dt}
b(t)=\exp{\{-itH\}}(-I)\exp{\{itH\}}
b(t),
\quad
b(0)=a_{{\bar k}{\bar M}_1{\bar M}_2\cdots}
\ene
in some approximation, and estimating the solution, as in [20],
 chapt. III, sect. 6, we see that
\beq
\nonumber
\sum_{\forall {M}_j}|b_{{\bar k}{M}_1{M}_2\cdots}(t)|^2
=1-\sum_{k\ne {\bar k},\ \forall M_j}|b_{{k}{ M}_1{ M}_2\cdots}(t)|^2
\ene
with the second term on the right-hand side
$\ge \mbox{\it const.} |t|$ ($\mbox{\it const.} \ne 0$)
 in norm for small $|t|$.
This, in turn by the definition above of $b_{{k}{ M}_1{ M}_2\cdots}(t)$,
 yields that there exists some energy shift $\delta E\ne 0$ to the
total energy of the system during the transition from the state $\bar k$
 to the same state $\bar k$ of the matter part of the system.
In this sense, it seems that the Lamb shift was already predicted
 at least implicitly around 1930.
We expect that this kind of argument could be refined in rigorous
 sense to give the explanation of the Lamb shift with no divergence
 problem, in our context of abandonment of relativistic quantum
field theory.

These conjectures stated here do not require us to introduce the notion
 of relativistic quantum field, because the system should be considered
as a non-relativistic one.
(See also Dirac [22], chapt. X other than [20], chapt. III.)
The relativistic quantum phenomena could be explained as the relativistic
 effects of observations as discussed in the section on observation.

\MP

\ \ \ 3) We adopt the standpoint that quantum mechanics is symmetric
 with respect to the reversal of (local) time: $t\to -t$.
 In this sense we stand upon the continuous
 $\langle\langle$Schr\"odinger-like$\rangle\rangle$ picture concerning
the measurement of the local systems by the observer's local system:
 As far as the local systems $(H_{n\ell},\HH_{n\ell})$ consisting of
finite number of particles are concerned, no bound state $\psi_{n\ell}$
of $H_{n\ell}$ in $\HH_{n\ell}$ with eigenvalue $\mu$ can be detected
if one observes the system in accord with the local time
$t=t_{(H_{n\ell},\HH_{n\ell})}$, because the state $\psi_{n\ell}$ does
 not change in its amplitude:
 $\vert e^{-itH_{n\ell}}\psi_{n\ell}(x)\vert^2=\vert
 e^{-it\mu}\psi_{n\ell}(x)\vert^2 =\vert \psi_{n\ell}(x)\vert^2$.
Therefore, it emits no light and information  outside.
Namely, only the scattering state which lies in the continuous spectral
 subspace $\HH^c_{n\ell}(\subset\HH_{n\ell})$ of the local Hamiltonian
 $H_{n\ell}$ can be detected.
 We need not consider here the so-called reduction of the wave packets.
 The observer measures only, {\it e.g.}, the light emitted from the
 scattering state in the observed system $(H_{n\ell},\HH_{n\ell})$.
In this sense the measurement does not include the irreversible procedure
in our theory, and the (local) time is reversible.
(We refer to von Neumann [20], chapt. III, VI for the arguments which
 support our standpoint including the explanation of emission of light
 in the line of conjecture (2).)

\MP

\ \ \ 4) By the consistency of {\it axioms} 4-5 with {\it axioms} 1-3,
 the theory of (general) relativity implies Hubble's red-shift.
Namely, the universe looks like it is expanding (or contracting) for any
 observer's local system.
At the same time, {\it axioms} 1-3 and {\it definitions} 1-3 imply that
 the universe has no proper time associated and is stationary.
This situation is no contradiction, because {\it axioms} 1-5 are consistent.
The expansion appears only when the observer's local system observes the
 universe in accordance with classical mechanics.
In other words, the universe looks like expanding only when the observer
 puts its concerns on the centres of mass of the stars, galaxies, etc.,
 regarding them as other local systems.
 Only in this context it is required to argue on the beginning and the end
 of the universe.
But an appearance of the universe as this comes from the viewpoint  merely
 based on classical mechanics.
When the observer looks at  the inside of those local systems, he sees that
 quantum mechanics is working inside them as we have seen in sect. 9.
Then he reasons from these observations and consideration that the
universe itself is a quantum-mechanical one, and concludes that the
consistent introduction of quantum mechanics into such a situation
would remove the problem of the beginning and the end of the universe.
His conclusion would be that the universe just
 $\langle\langle$exists$\rangle\rangle$ without time.

\BP


\small


\begin{thebibliography}{98}

\vskip 8pt

\bibitem{2}
 A. Ashtekar, J. Stachel (eds.),
{\it Conceptual Problems of Quantum Gravity},
 Birkh\"auser, Boston-Basel-Berlin,
 1991.

\bibitem{3} H.R. Brown, R. Harr\' e (eds.),
{\it Philosophical Foundations of Quantum Field Theory},
 Clarendon Press, Oxford,
 1990.



\bibitem{11}
 J. Glimm, A. Jaffe
{\it Quantum Physics, A Functional Integration Point of View},
 Springer-Verlag, 2nd edn.,
New York-Berlin-Heidelberg-London-Paris-Tokyo,
 1987.


\bibitem{19} M. Redhead, {\it Incompleteness, Nonlocality, and Realism,
A Prolegomenon to the Philosophy of Quantum Mechanics},
 Clarendon Press, Oxford, 1990.


\bibitem{18}
 E. Prugove\v cki,
{\it Quantum Geometry, A Framework for Quantum General Relativity},
 Kluwer Academic Publishers, Dordrecht-Boston-London,
 1992.







\bibitem{1}R. Abraham, J.E. Marsden,   {\it Foundations of Mechanics},
 The Benjamin/Cummings  Publishing  Company, 2nd edn.,
London-Amsterdam-Don Mills, Ontario-Sydney-Tokyo,
 1978.



\bibitem{15} H. Kitada, {\it Asymptotic completeness of N-body wave
operators  I. Short-range quantum systems}, Rev. Math. Phys. {\bf 3}, 1991,
101-124.


\bibitem{6}V. Enss,
{\it Introduction  to  asymptotic
observables  for  multiparticle quantum scattering}, in \lq\lq
Schr\"odinger
Operators, Aarhus 1985,'' ed. E. Balslev,
 Lect.  Note in  Math. {\bf 1218},
Springer-Verlag,  1986,
pp.61-92.



\bibitem{16} H. Kitada, {\it Asymptotic completeness of N-body wave
operators  II. A new proof for the short-range case and the asymptotic
clustering for long-range systems},
in \lq\lq Functional Analysis and Related Topics, 1991,'' ed. H. Komatsu,
 Lect. Note in Math. {\bf 1540}, Springer-Verlag, 1993, pp.149-189.



\bibitem{12} S.W. Hawking, G.F.R. Ellis, {\it The Large Scale Structure of
Space-Time}, Cambridge University Press, Cambridge-New York-Port
Chester-Melbourne-Sydney, 1973.


\bibitem{4} J. Derezi\' nski, {\it Asymptotic completeness of
long-range N-body quantum systems},
 preprint (Centre de Math\'ematiques, Ecole Polytechnique, U.R.A.
169 du C.N.R.S.), 1992.


\bibitem{7}A. Einstein, {\it Die Grundlage der allgemeinen
Relativit\" atstheorie},   Ann. der Phys. (Leibzig), Ser. 4,
{\bf 49}, 1916, 769-822.



\bibitem{10}  M. Friedman, {\it Foundations of Space-Time Theories,
Relativistic Physics and Philosophy of Science},
 Princeton University Press, Princeton, 1983.


\bibitem{8} A. Einstein, B. Podolsky, N. Rosen, {\it Can quantum-mechanical
description of physical reality be considered complete} ?,
 Phys. Rev. {\bf 47},  1935, 777-780.



\bibitem{13} M. Jammer, {\it The Philosophy of Quantum Mechanics, The
interpretations of Quantum Mechanics in Historical Perspective},
John Wiley \& Sons, Inc., New York, 1974.


\bibitem{20} F. Selleri, {\it Die Debatte um die Quantentheorie}, Frieder.
Vieweg \& Sohn Verlagsgesellschaft mbH, Braunschweig, 1983.



\bibitem{21}  W. Schommers (ed.), {\it Quantum Theory and Pictures of Reality,
Foundations, Interpretations, and New Aspects},
 (with contributions by B. d'Espagnat, P. Eberhard, W. Schommers, F.
Selleri),
 Springer-Verlag, Berlin-Heidelberg-New York-London-Paris-Tokyo-Hong
Kong, 1989.



\bibitem{17} N.F. Mott, H.S.W. Massey,
{\it The Theory of Atomic Collisions},
 Clarendon Press, 3rd edn., Oxford,
 1965.


\bibitem{14} A. Jensen, T. Kato, {\it Spectral properties of Schr\" odinger
operators and time-decay of the wave functions}, Duke Math. J.
{\bf 46},  1979,  583-611.



\bibitem{22}J. von Neumann, {\it Die Mathematische Grundlagen der
Quantenmechanik},
  Springer-Verlag, Berlin, 1932.



\bibitem{9} R.P. Feynman, A.R. Hibbs,
{\it Quantum Mechanics and Path Integrals}, McGraw-Hill Book Company,
New York-St. Louis-San Francisco-Toronto-London-Sydney, 1965.







\bibitem{5}
 P.A.M. Dirac,
{\it The Principles of Quantum Mechanics},
 Clarendon Press, 4th edn., Oxford,
 1958.




\end{thebibliography}
\end{document}